\documentclass[
  ,final            % use final for the camera ready runs
    ,draft            % use draft while you are working on the paper
  ,numberedheadings % uncomment this option for numbered sections
  ]
  {aipproc}

\layoutstyle{8x11single}

\begin{document}

\title{Trigger Study for KM3Net-Italia}

\classification{07.05.Hd, 95.55.Vj}
\keywords      {Data acquisition, neutrino telescopes, muon triggers}

\author{B. Bouhadef$^1$\footnote{Corresponding author (email: bouhadef@pi.infn.it)}}{
  address={INFN Sezione di Pisa, Italy},
 ,email={bouhadef@pi.infn.it}
}

\author{the NEMO collaboration}{
   address={\url{http://www.pi.infn.it/~bouhadef/authors_list}}
   }

\begin{abstract}
A new trigger for NEMO Phase 2 tower based on the time differences of the PMT hits has been studied. Such a trigger uses only a fixed number of PMT hits in a chosen time windows. The background trigger rate is drastically reduced requiring hits from different PMTs. A 87\% trigger efficiency was estimated by Montecarlo simulation for muon tracks with at least 5 PMT hits. The trigger rate estimated by Montecarlo was also measured on raw data. The results from Montecarlo simulations and raw data are reported.
\end{abstract}

\maketitle

%%%%%%%%%%%%%%%%%%%%%%%%%%%%%%%%%%%%%%%%%%%%
%% MAINMATTER
%%%%%%%%%%%%%%%%%%%%%%%%%%%%%%%%%%%%%%%%%%%%

\section{Introduction}

		A Neutrino telescope is a tool used to increase our knowledge and to answer fundamental questions about the Universe. Following the success of the IceCube experiment ~\cite{ICECUBE0}, which is a km3 size telescope in the ice at south pole, and of the ANTARES experiment ~\cite{ANTARES0}, an underwater telescope with a volume of 0.4km$^3$, the European scientific community is going to construct a neutrino telescope similar to but larger than IceCube called Km3Net in the Mediterranean Sea. New detector prototypes for KM3 were also studied by NEMO (phase I and II) ~\cite{NEMO0}  and NESTOR  ~\cite{NESTOR0}.
	\smallskip	 		 
		
		The NEMO collaboration in May 2013 have deployed a tower of 8 floors at Capo Passero site, each floor equipped with 2 pairs of PMTs as shown in Fig. 1 (NEMO phase II). During the data taking of NEMO phase II, the average measured rates of 31/32 PMTs surviving the deployment were about $50 kHz$, which means that we measured a total PMTs rate less than $2 MHz$ with respect to the expected few Hz  of atmospheric muon tracks, and about few micro Hz up-going muons. In such situations, the muon tracks tagging is a challenge:  the trigger must be efficient and be able to reduce the optical background.
		\smallskip	     
 
	Up to know all underwater telescopes are using the time and the charge information to select a candidate muon tracks and to reduce the background rate, our proposed trigger uses only the time information of N consecutive hits within a maximum fixed Time Windows (TW) before using time coincidences criteria. Such a trigger is immune to charge calibration errors. Moreover, the charge information can be used at high level trigger.

	In such a trigger, the number of hits N as well as TW can be adjusted to achieve the desired trigger background rate. The next two sections deal with the study of the background rates and the trigger efficiency for the muon track tagging.

%==========================================================================
%    Fig 1
%==========================================================================
%\newpage
	 
	\begin{figure}
  \includegraphics[height=.3\textheight]{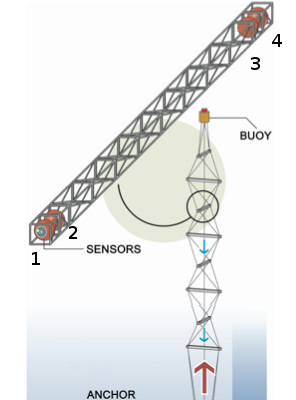} \\
  \includegraphics[height=.3\textheight]{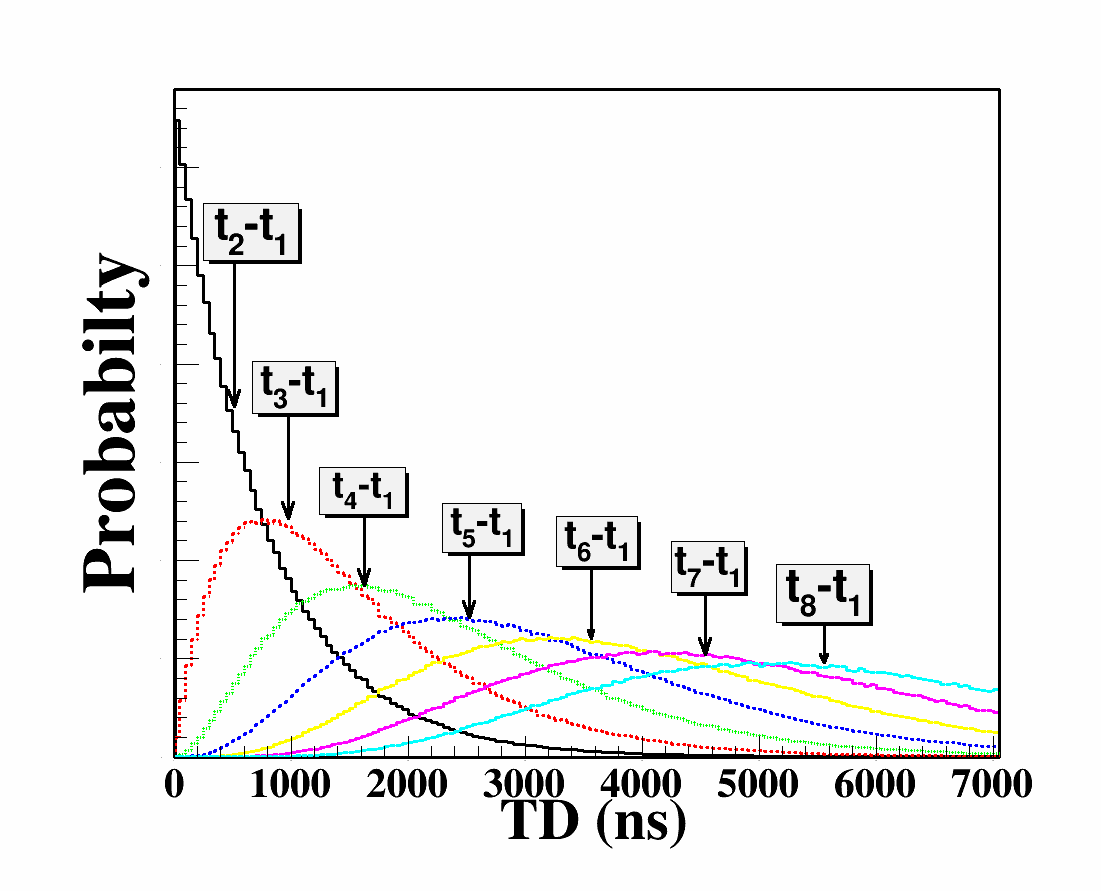} \\
  \caption{Left: Nemo Tower Phase II. Right: Time Difference distributions between the first hit and 2$^{end}$,.., 8$^{th}$ background hit. }\label{Fig:1}
\end{figure}

%====================================================================================
%   Fig.2 for 47 and 56.7 triggering for DN6, 7 and 8
%====================================================================================

	\begin{figure}
  \includegraphics[height=.25\textheight]{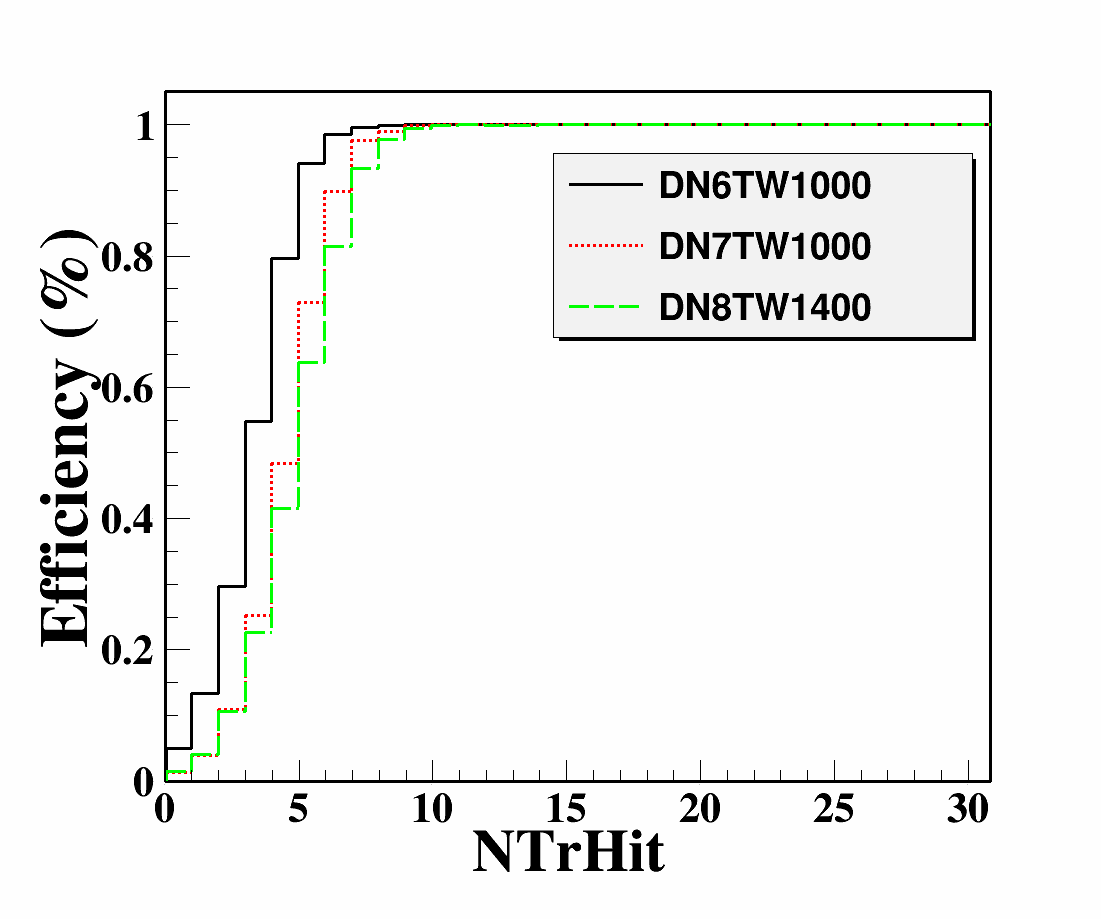} \\
   \includegraphics[height=.25\textheight]{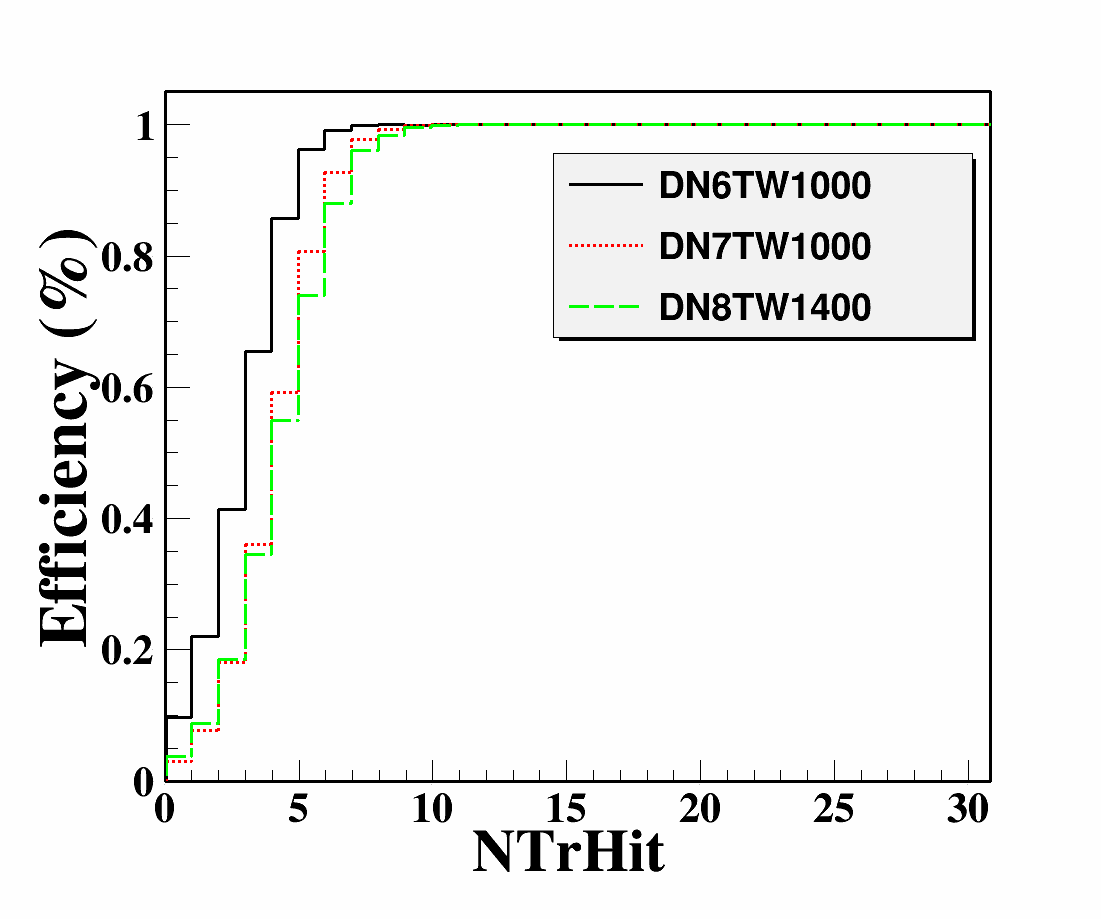} \\
  \caption{Trigger efficiency of DN6TW1000 (black),DN7TW1000 (red) and DN8TW1400 yellow @ a background rate s 47kHz(left) and 57kHz(right).}
\end{figure}\label{Fig:2}

\section{Trigger Efficiency Study}

	To study the muon track tagging efficiency we have used a Montecarlo based on Mupage ~\cite{MUPAGEPAC} and KM3 ~\cite{KM3PAC} packages
for atmospheric muon generation and propagation inside the detector. After that, we have added each muon track (with NtrHit>=1) hits to 10 $\mu s$ of background and we have applied the electronic front-end simulation of NEMO phase II. The background extraction was repeated 100 time for each muon track. 

	The generated atmospheric muon tracks are distributed between  $0-85^{\circ}$ along the zenith angle, and we have verified that the arrival time of direct photon hits are within 1$\mu s$, whereas the hits of scattered photons increase the arrival times windows to more than 1.4 $\mu s$. For this reason, we have chosen the TWs to be in the 1-1.4 $\mu s$ range (TW1000 - TW1400).
	
Fig. 1 (right) shows the TD distributions @ $47.2 kHz$ background rate  between first hit time and 2$^{end}$,3$^{th}$,.., 8$^{th}$ hit time of the 8 different PMTs, and we observe that the mean TD distribution increases from 2 to 8 background hits. This means that the probability of finding N background hits (NBkHit) in a fixed TW decreases as the requested N hits increases.% the NBkHit increases. 
\smallskip	
In Table 1, we report the conditional probabilities to have only N (1-8) background hits in different TWs. In addition, the probability to have N>=7 and N>=8 are also reported. The probability to tag a muon track with 5 PMT hits in TW1000 depends on the probablity to find in the same TW a number of background hits >= 2, and is  not more than 75\% requiring 7 different PMT hits (DN7), and it is less than 54\% (background hits >= 3) in TW1400 requiring DN8.
	
\smallskip	
	
Fig. 2 left (right) shows at $47.2 kHz$  ($56.7 kHz$) background rate the trigger efficiencies requiring 6 hits from Different PMTs in the TW1000 (DN6TW1000 trigger: solid black line), 7 hits from Different PMTs in the TW1200 (DN7TW1200 trigger: dotted red line), and 8 hits from Different PMTs in the TW1400 (DN8TW1400 trigger: dashed green line). The trigger efficiency as function of muon hit number is calculated dividing the number of the TW triggers in which there is at least one muon track hit by the number of the background extractions (if a muon track triggers few time  in a single extraction is considered as a single trigger).  A trigger efficiency of more than 80\% is achieved with NTrHit>=8 and the trigger rates range from $0.2 kHz$ in the DN8TW1000 @ $47.2 kHz$ case to about $30 kHz$ in the DN6TW1400 @ $56.7 kHz$ case.

%=====================================================================================
%           Table 1,2 for conditional probabilities and corresponding rates
%=====================================================================================
\begin{center}

\begin{table}
\begin{tabular}{lrrrrrrrrrr}
\hline
  & \tablehead{1}{r}{b}{N1}
  & \tablehead{1}{r}{b}{N2}
  & \tablehead{1}{r}{b}{N3}
  & \tablehead{1}{r}{b}{N4}   
  & \tablehead{1}{r}{b}{N5} 
  & \tablehead{1}{r}{b}{N6} 
  & \tablehead{1}{r}{b}{N7}
  & \tablehead{1}{r}{b}{N8}
  & \tablehead{1}{r}{b}{N>=7}
  & \tablehead{1}{r}{b}{N>=8} \\
\hline
TW1000 & 25\% & 36\% & 23.5\% & 10.5\% & 3.8\% & 0.8\% & 0.16\% & 0.03\% &0.23\% & 0.04\%\\
TW1200 & 19\% & 35\% & 33\% & 14\% & 5.6\% & 1.6\% & 0.37\% & 0.08\% &0.47\% & 0.09\%\\
TW1400 & 14\% & 29\% & 28\% & 17\% & 7.8\% & 2.7\% & 0.74\% & 0.12\% &0.90\% & 0.15\%\\
\hline
\end{tabular}
\caption{Conditional probabilities of N hits in different TWs. The last two colums are the probability to have N>=7,8 hits in the TWs.}
\label{tab:1}
\end{table}

\bigskip

\end{center}

\begin{table}
\begin{tabular}{lrrrrrrrrrr}
\hline
  & \tablehead{1}{r}{b}{N1}
  & \tablehead{1}{r}{b}{N2}
  & \tablehead{1}{r}{b}{N3}
  & \tablehead{1}{r}{b}{N4}   
  & \tablehead{1}{r}{b}{N5} 
  & \tablehead{1}{r}{b}{N6} 
  & \tablehead{1}{r}{b}{N7}
  & \tablehead{1}{r}{b}{N8}
  & \tablehead{1}{r}{b}{N>=7}
  & \tablehead{1}{r}{b}{N>=8} \\
\hline
TW1000 & 365 & 518 & 350 & 156 & 55.5 & 12.3 & 2.4 & 0.38 & 2.8 & 0.83 \\
TW1200 & 277 & 501 & 474 & 207 & 82.0 & 23.0 & 5.4 & 1.1 & 6.8 & 1.3 \\
TW1400 & 210 & 424 & 409 & 254 & 113 & 37.4 & 10.9 & 1.7 & 13.1 & 2.2 \\
\hline
\end{tabular}
\caption{The expected rates  in kHz of selecting N hits in different TWs @ background Rate of $47.2 kHz$.}
\label{tab:2}
\end{table}

\bigskip

\begin{table}
\begin{tabular}{lrrrrrrrrrr}
\hline
  & \tablehead{1}{r}{b}{muon tracks}
  & \tablehead{1}{r}{b}{N7TW1000}
  & \tablehead{1}{r}{b}{SCCut}
  & \tablehead{1}{r}{b}{FCCut}
  & \tablehead{1}{r}{b}{DN5Cut}  \\ 
\hline
 Rates(Hz)\textbf{(NTrHIT>=1) }& \textbf{5.43} & 5100/\textbf{1.18} & 800/\textbf{0.79} & 400/\textbf{0.73} & 250/\textbf{0.73}  \\
 Rates(Hz)\textbf{(NTrHIT>=5)} & \textbf{0.88 }& 5100/\textbf{0.77} & 800/\textbf{0.63} & 400/\textbf{0.62} & 250/\textbf{0.62} \\
\hline
\hline
  & \tablehead{1}{r}{b}{muon tracks}
  & \tablehead{1}{r}{b}{DN7TW1000}
  & \tablehead{1}{r}{b}{SCCut}
  & \tablehead{1}{r}{b}{FCCut}
  & \tablehead{1}{r}{b}{DN5Cut}  \\ 
\hline
 Rates(Hz)\textbf{(NTrHIT>=1) }& \textbf{5.43} & 3000/\textbf{0.97} & 500/\textbf{0.69} & 200/\textbf{0.65} & 200/\textbf{0.65}  \\
 Rates(Hz)\textbf{(NTrHIT>=5)} & \textbf{0.88 }& 3000/\textbf{0.71} & 500/\textbf{0.59} & 200/\textbf{0.58} & 200/\textbf{0.58} \\
\hline
\end{tabular}
\caption{The expected rates  in Hz of selecting N (DN) hits in different TWs @ background Rate of $47.2 kHz$.}
\label{tab:3}
\end{table}

	Looking at Table 2 and Fig. 2, a good choice is DN7TW1000 trigger with a background rate less than 3 kHz (N>=7) @ $47.2 kHz$. Also, considering 7 different hits searching in the TW1000 is time consuming, we propose to tag 7 hits in TW1000s also coming from same PMT (N7TW1000 trigger instead of the DN7TW1000 trigger). The TWs triggered by the N7TW1000 contain all TWs triggered by DN7TW1000 with an increasing of the background rate, which means that N7TW1000 is more efficient than DN7TW1000.

%====================================================================================
%   Fig.4 for N7TW1000 and muon track rates
%====================================================================================

	\begin{figure}
  \includegraphics[height=.25\textheight]{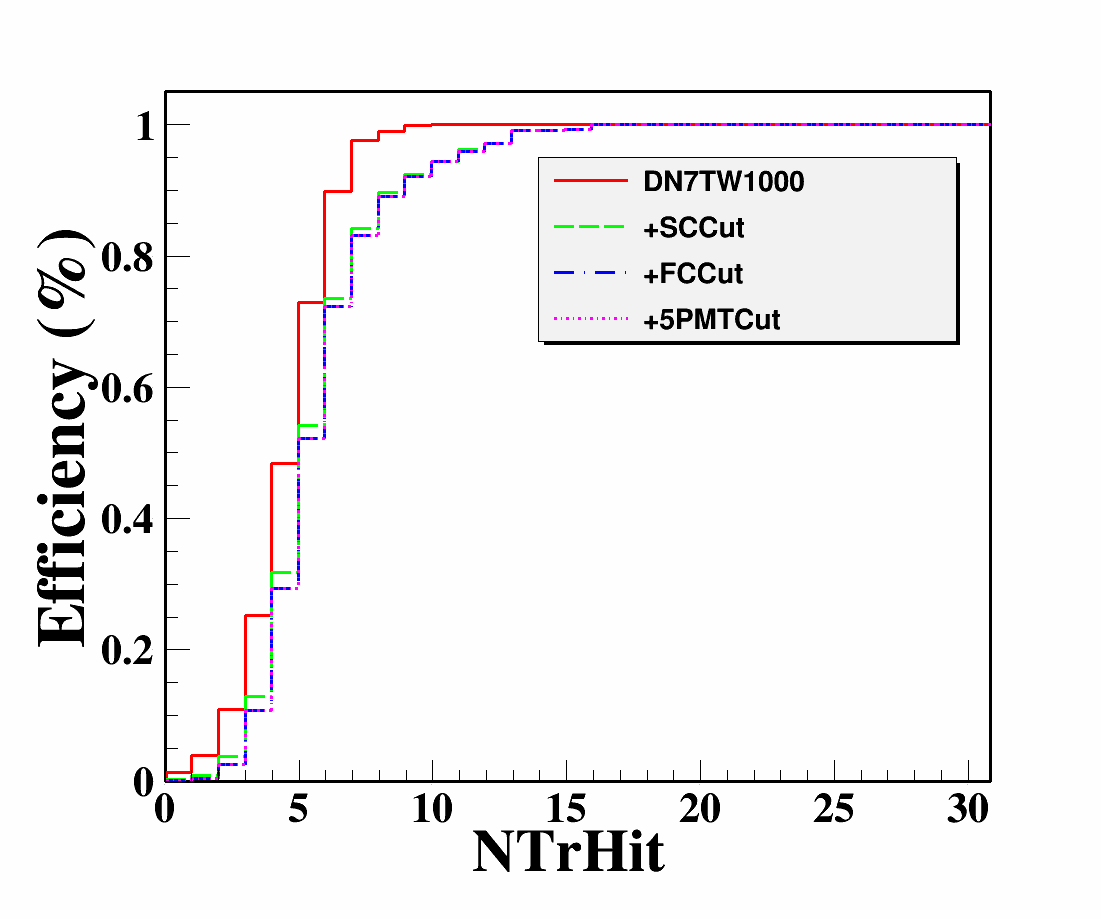} \\
   \includegraphics[height=.25\textheight]{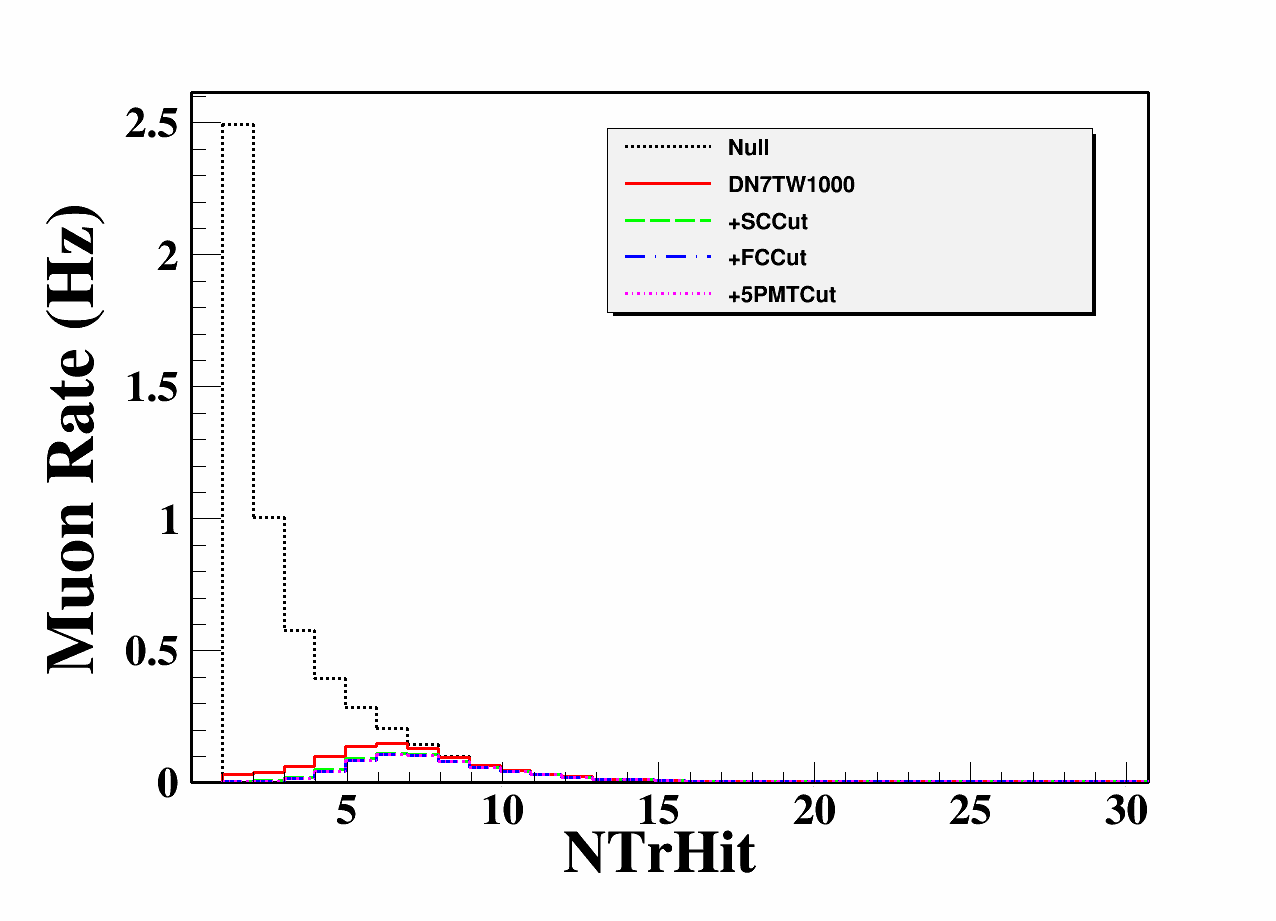} \\
  \caption{Trigger efficiency of DN7TW1000 different cuts (left), and the corresponding expected atmospheric muons rates.}\label{Fig:3}
\end{figure}

%==================================================================================
%      Table 4 for mc and real data confrontation
%==================================================================================

\begin{table}
\begin{tabular}{lrrrrrrrrrr}
\hline

  & \tablehead{1}{r}{b}{N7TW1000}
  & \tablehead{1}{r}{b}{SCCut}
  & \tablehead{1}{r}{b}{FCCut}
  & \tablehead{1}{r}{b}{DN5Cut}  \\ 
  \hline
   MC/raw data & 5100/5200 & 800/840 & 400/330 & 250/200 \\
\hline
 & \tablehead{1}{r}{b}{DN7TW1000}
  & \tablehead{1}{r}{b}{SCCut}
  & \tablehead{1}{r}{b}{FCCut}
  & \tablehead{1}{r}{b}{DN5Cut}  \\ 
  \hline
   MC/raw data & 3000/3000 & 400/720 & 200/190 & 200/150 \\
\hline
\end{tabular}
\caption{The expected rates  in Hz form Montecarlo and raw data @ background Rate of 47.2kHz.}
\label{tab:4}
\end{table}

	In the Fig. 3 and the table 3  are reported the N7TW1000 trigger efficiencies and the tagged muon track frequency as calculated with the Montecarlo simulation; in the Table 3 are also reported the corresponding estimated trigger rates. We observe that the estimated rate increases to about $5 kHz$ compared to DN7TW1000 trigger, and with a better muon tagging efficiency. In conclusion using the N7TW1000 (7 hits in TW=1000 also from same PMT) trigger increases the total efficiency from 80\% to 87\% with NtrHit>=5.

\section{Results and Conclusion}

Starting from N7TW1000 as level 1 trigger (requiring 7 hits in a Time Window of maximum 1000 ns) the calculated trigger rate is $5.1 kHz$ @ $47 kHz$ per PMT (see Table 4) and the measured raw data trigger rate was measured to be nearly the same. By adding other conditions such two PMTs coincidence in same floor (SCCut) and two PMTs coincidence in adjacent floors (FCCut) the rate is reduced by factor 10. By requiring 5 hits in different PMTs (DN5Cut), we reach a trigger rate of about 200 Hz. The bold numbers in the Table 3 are the muon track rates tagged by each sequential cut. 
\smallskip	

The level 2 trigger cuts also reduce the N7TW1000 sensitivity to the bio-luminescences as we have seen on real data (not shown here), particularly, the DN5Cut is used to reduce the bio-luminescence effect and to ensure 5 different PMT hits for muon track reconstruction. In a simulation (not shown here) we find that the N7TW1000 is an equivalent trigger to the PMTs coincidence trigger (in same floor) with a reduced background about 40\%. Moreover using TW trigger before SCCut reduces the trigger computations, because the SCCut (which is a necessary condition) will not be applied on all data. 

\smallskip	

	The efficiencies of N7TW1000 and the other cuts are shown in Fig. 3 (left), and their estimated muon rates (right). The graph with dotted black color is the total estimated muon rate with an integral rate of $5.4 Hz$.  The successive trigger cuts were also applied on raw data. The total muon rates are reported in Table 3. The DN5Cut trigger has the same rate and same efficiency to FCCut, and is used in presence of bio-luminescences.
	
\smallskip	
	      
	Next step of KM3NeT will be the construction of 8 towers similar to the NEMO phase II prototype to be deployed in the 2015 in Capo Passero site: in the final towers,  the inter-floor distance will be reduced to 20m and the number of PMTs by floor will be increased to 6 (instead of 4 PMTs). A preliminary study of the TW trigger applied to the new tower design shows that the expected rate can be reduced to less than 200 Hz/tower without losing efficiency for NTrHit>=5/tower.

%%%%%%%%%%%%%%%%%%%%%%%%%%%%%%%%%%%%%%%%%%%%%%%%%%%%% end paper  %%%%%%%%%%%%%%%%%%%%%%%%%%%%%%%%%%%%%%%%%%%%%%%%%%%%%%%

%%%%%%%%%%%%%%%%%%%%%%%%%%%%%%%%%%%%%%%%%%%%%%%%
%% BACKMATTER
%%%%%%%%%%%%%%%%%%%%%%%%%%%%%%%%%%%%%%%%%%%%%%%%

%\begin{theacknowledgments}
%\end{theacknowledgments}



\bibliographystyle{aipproc}   % if natbib is available

\bibliography{vlvnt13}

\IfFileExists{\jobname.bbl}{}
 {\typeout{}
  \typeout{******************************************}
  \typeout{** Please run "bibtex \jobname" to optain}
  \typeout{** the bibliography and then re-run LaTeX}
  \typeout{** twice to fix the references!}
  \typeout{******************************************}
  \typeout{}
 }

\end{document}